# Identifying Cheating Anchor Nodes using Maximum Likelihood and Mahalanobis Distance

Jeril Kuriakose[A], Amruth V[B]  Swathy Nandhini[C]  Abhilash V[D]

[A]School of Computing and Information Technology (SCIT), Manipal University Jaipur, Jaipur, India, Contact: jeril@muj.manipal.edu

[B]Department of Information Science and Engineering, Bearys Institute of Technology, Mangalore, India.

[C]Department of Information Technology, Jayam College of Engineering and Technology, Dharmapuri, India.

[D]Freelancer.

Malicious anchor nodes will constantly hinder genuine and appropriate localization. Discovering the malicious or vulnerable anchor node is an essential problem in Wireless Sensor Networks (WSNs). In wireless sensor networks, anchor nodes are the nodes that know its current location. Neighbouring nodes or non-anchor nodes calculate its location (or its location reference) with the help of anchor nodes. Ingenuous localization is not possible in the presence of a cheating anchor node or a cheating node. Nowadays, it's a challenging task to identify the cheating anchor node or cheating node in a network. Even after finding out the location of the cheating anchor node, there is no assurance, that the identified node is legitimate or not. This paper aims to localize the cheating anchor nodes using trilateration algorithm and later associate it with maximum likelihood expectation technique (MLE), and Mahalanobis distance to obtain maximum accuracy in identifying malicious or cheating anchor nodes during localization. We were able to attain a considerable reduction in the error achieved during localization. For implementation purpose we simulated our scheme using ns-3 network simulator.

## 1. INTRODUCTION

Wireless Adhoc and sensor networks are on a steady rise in the recent decade. This is because of their reduced cost in deployment and maintenance. Advancements in radio frequency spectrum also carved way for the improvement in the data rate for communication. Many devices belong to wireless ad hoc and sensor networks; one among them is anchor node [1-8]. Anchor nodes are the nodes that know its current location. Neighbouring nodes or non-anchor nodes calculate its location (or location reference) with the help of an-chor nodes, and its working is quite referable to Light House.

The location of the nodes plays a significant role in many areas as routing, surveillance and monitoring, military etc. The sensor nodes must know their location reference to carry-out location-based routing (LR) [9-12]. To find out the shortest route, the location aided routing (LAR) [13-15] makes use of the locality reference of the sensor nodes. In some industries the sensor nodes are used to identify minute changes as pressure, temperature and gas leak, and in military, robots are used to detect land-mine where in both the cases location informa-tion plays a key part.

Anchor nodes can also be used to find the current location of any device (mobile phones, ob-jects and people). It does that by transmitting anchor frames periodically or at regular intervals. Usually anchor frames are used to advertise the occurrence of a wireless modem or an Access Point (AP). Each anchor frame carries some details about the configuration of AP and a little security information for the clients.

When the technologies are on a massive upswing, the need for security of the relevant technologies arises. There can be several occasion where the anchor nodes can be vulnerable to security breach. Because of the security breach the anchor node starts cheating by providing false information. In the pres-ence of cheating anchor nodes the chances of localization drastically decreases. Many pa-pers [16-19] discuss about the localization of cheating anchor nodes, but with inconsistent accuracy. So, to overcome this, we localize the cheating or vulnerable anchor node using trilateration technique, and associate the results with maximum likelihood expectation tech-nique [20-33] and Mahalanobis distance [34]. No such scheme has been used till now to iden-tify the malicious anchor nodes.

Organization of the paper: Section 2 pro-vides the localization using trilateration algo-rithm and Section 3 and Section 4 studies the maximum likelihood expectation and Mahalanobis distance, respectively. Simulation and results are covered in Section 5, Section 6 expli-cates few future events and Section 7 concludes the paper.

## 2. LOCALIZATION USING TRILATERATION ALGORITHM

Anchor nodes are widely used for tacking and localization; whereas now-a-days it is also used for navigation and route-identification. With the help of anchor nodes, a user can find out his current location. Consider a scenario like a hotel or museum, there may be many occasions where people go out of track. This can be flab-bergasted by installing anchor nodes installed in various locations, so that people can trace out there location very easily and it is possi-ble only when the anchor nodes are authen-tic. Now-a-days hackers are on a rise; anybody can easily get into any system and change its settings. Similarly, they can hack any anchor nodes and change its location reference to some other false location reference, making people lose their track; thus leading to a bad imprint about the system (i.e., hotel, museum).

An attack is exemplified in Figure 1 and Figure 2. Figure 1 shows the initial deployment of anchor nodes $A_1$, $A_2$, $A_3$; with location reference $(x_1, y_1)$, $(x_2, y_2)$, $(x_3, y_3)$; and distance $L_1$, $L_2$, $L_3$; respectively, from the trilateration point T, having location reference $(x_t, y_t)$. Figure 2 demonstrates the logical deployment of anchor nodes after the attack i.e., multiple changes in location reference of anchor node $A_2$.

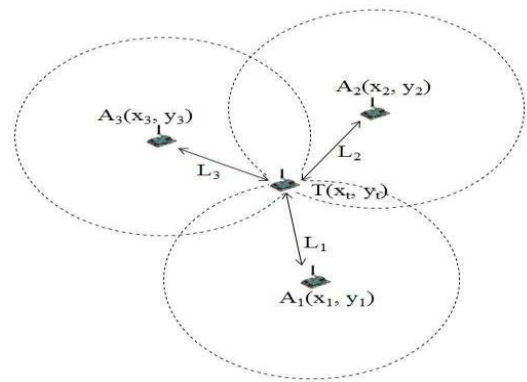

Figure 1. Initial Set-up of Anchor Nodes.

The three dimensional location coordinate of any device or node can be estimated using trilateration calculations. Trilateration technique uses distance measurements rather than angular measurements; latter technique is also used in many localization techniques [21-23]. Us-ing some iterative schemes like least square, least median square [17], least trimmed square [24] and gradient descent [25], can equitably increase the accuracy of trilateration technique.

Trilateration techniques use the distance measurement between the nodes to calculate the location reference. The distances between the nodes are identified using Received Sig-nal Strength (RSSI) [26, 33] or Time of Ar-rival (ToA) [27, 28, 33] or Time Difference of Arrival (TDoA) [29, 30, 33]. When a node (re-questing node) wants to identify its location in-formation using trilateration technique, it does with the help of three or more neighbouring anchor nodes. The exemplification of trilater-ation techniques is as follows:

1. A node that wants to find its location reference (or location coordinate) sends a localization request to any of its neighbouring anchor nodes. The anchor node sends a reply with its current location reference and its RSSI measurement with respect to the node that wants to localize. Based on this information, we put up a virtual wireless ring (VWR) (or logical ring) [31] as shown in Figure 3. The assumption of the logical ring is made with the an-chor node as centre. The requesting node can be located anywhere on the circum-ference of the logical ring, and thus mak-ing it difficult to guess its exact location.

2. Next the same requesting node sends another localization request to a differ-ent neighbouring anchor node. The an-chor node follows the same process as discussed in the previous step. Again another logical ring is updated to the pre-vious one, shown in Figure 4. From the logical observation we can analyse that the location of the requesting node could be present in any one of the intersecting point of the two logical rings.

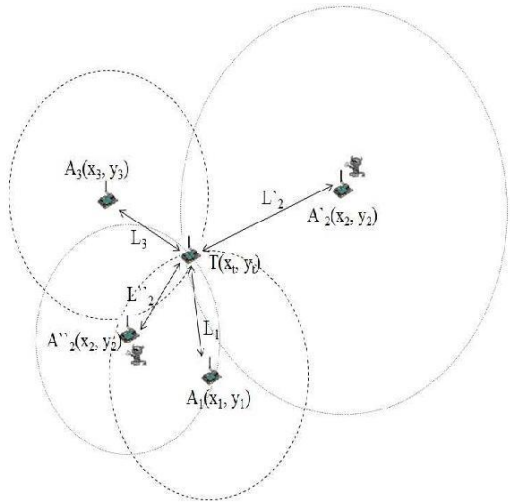

Figure 2. Initial Set-up of Anchor Nodes.

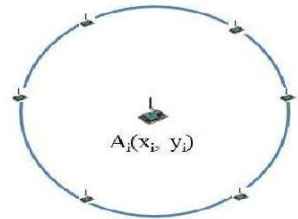

Figure 3. Virtual wireless ring with one anchor node.

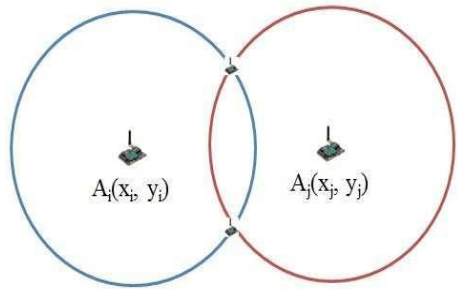

Figure 4. Virtual Wireless Ring with Two An-chor Nodes.

3. Finally to ease the muddle, the same re-questing node sends another localization request to a different neighbouring an-chor node other than the previous two anchor nodes. The same process is re-peated with the new neighbouring anchor node. When the final virtual wireless ring is drawn, we would be able to extract the exact location of the requesting node. Figure 5 shows the localization of a node using trilateration technique.

The three dimensional location coordinate of any device or node can be estimated using tri-lateration calculations. Trilateration technique uses distance measurements rather than angu-lar measurements; the latter technique is also used in many localization techniques. Using some iterative schemes like least square, least median square, least trimmed square and gra-dient descent, can equitably increase the accu-racy of trilateration technique.

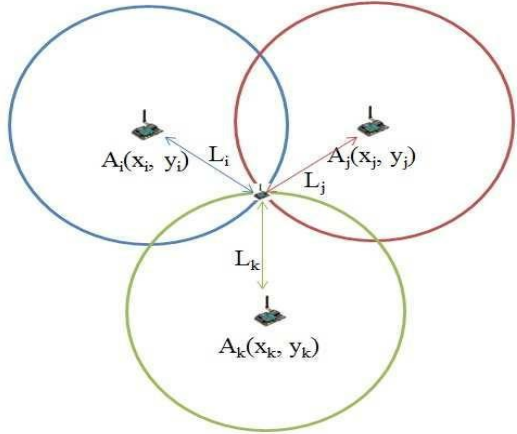

Figure 5. Virtual Wireless Ring with Three Anchor Node.

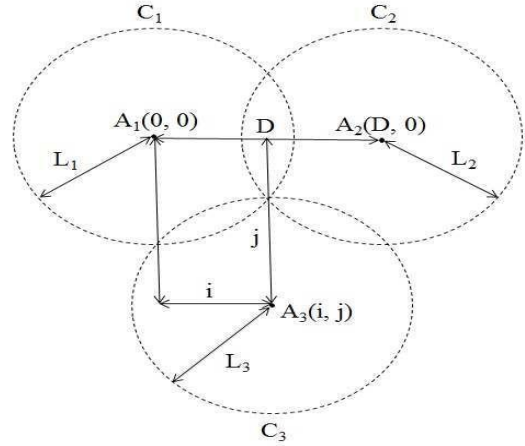

Figure 6. Trilateration Measurements.

Trilateration techniques use the distance mea-surement between the nodes to calculate the location reference. The distances between the nodes are identified using Received Signal Strength (RSSI) or Time of Arrival (ToA) or Time Difference of Arrival (TDoA). When a node (requesting node) wants to identify its location information using trilateration tech-nique, it does with the help of three or more neighbouring anchor nodes.

The mathematical computation of trilateration is as follows:
Consider three circles or spheres with centre $C_1$, $C_2$ and $C_3$, radius $L_1$, $L_2$ and $L_3$ from points $A_1$, $A_2$ and $A_3$ (anchor node location),
refer Figure 6.
The general equation of the sphere is

$$\sum_{k=1}^{3} (A_k - C_k)^2 = L^2$$

This can be modified as follows,

$$L_1^2 = A_1^2 + A_2^2 + A_3^2 \quad (1)$$

$$L_2^2 = (A_1 - D)^2 + A_2^2 + A_3^2 \quad (2)$$

$$L_3^2 = (A_1 - i)^2 + (A_2 - j)^2 + A_3^2 \quad (3)$$

Subtracting Eq. (2) from Eq. (1), we get
$$L_2^2 - L_1^2 = (A_1 - D)^2 + A_2^2 + A_3^2 - A_1^2 - A_2^2 - A_3^2 \quad (4)$$

Substituting we get,

$$A_1 = \frac{L_1^2 - L_2^2 + D^2}{2D} \quad (5)$$

From the first two circles we can find out that the two circles intersect at two different points, that is

$$D - A_1 < A_2 < D + A_1 \quad (6)$$

Substituting Eq. (5) in Eq.(1), we can procure

$$L_1^2 = \left(\frac{L_1^2 - L_2^2 + D^2}{2D}\right)^2 + A_2^2 + A_3^2 \quad (7)$$

Substituting we get the solution of the inter-section of two circles

$$A_2^2 + A_3^2 = L_1^2 - \frac{(L_1^2 - L_2^2 + D^2)^2}{4D^2} \quad (8)$$

Substituting Eq. (1) with Eqs. (3) and (8), we get
$$s = L_1^2 - A_1^2 - A_2^2$$

$$L_3^2 = (A_1 - i)^2 + (A_2 - j)^2 + s \quad (9)$$

$$A_2 = \frac{L_1^2 - L_2^2 - A_1^2 + (A_1 - i)^2 + j^2}{2j}$$

$$= \frac{L_1^2 - L_2^2 + i^2 + j^2}{2j}$$

$$A_2 = \frac{i}{j} L_1 \quad (10)$$

From Eq. (5) and Eq. (10) we get the values of $A_1$ and $A_2$ respectively. From that we can find out the value of $A_3$ from Eq. (1),

$$A_3 = \pm \sqrt{L_1^2 - A_1^2 - A_2^2}$$

From the above equation we can say that, $A_3$ can have either positive or negative value. If any one circle intersect the other two circles precisely at one point, then $A_3$ will get a value zero. If it intersects at two or more points, outside or inside it can get either a positive or negative value, respectively.

During deployment each node carries out the trilateration process with all of its neighbour-ing nodes and every node is authorized with two or more trilateration points for security reasons. Every node reveals the information about its trilateration point to its immediate or one hop neighbours. Care is taken that no node reveals the trilateration information about its neighbours.

The algorithm for setting up the anchor nodes according to trilateration is given in algorithm 1:

The algorithm for finding out the malicious anchor nodes is shown in algorithm 2:

After the comparison, the anchor nodes that does not have the same location reference or the anchor node that tends to be vulnerable is considered to be malicious or cheating node. To confirm its adversary, we compare it with maximum likelihood expectation and Maha-lanobis distance.

Data: Deployment of anchor nodes
Result: Successfully deploying anchor nodes and exchanging trilateration information

Start initialization; Deploy the anchor nodes;
Set the initial coordinates (lat & long) for each anchor node;
Cluster anchor nodes into a set of three or more;
while not at end of deployment do
    Trilaterate a group of anchor nodes to a centre point (or trilateration point) and save the location reference in $M_1^*$;
    Individually trilaterate all the anchor nodes with the neighbouring group and save the location references in $M_2^*$, $M_3^*$, etc.;
    Pass trilateration information to its immediate neighbours;
end
Stop deploying anchor nodes;
(* $M_1$, $M_2$, $M_3$, ETC., are different memory with different location reference)

Algorithm 1: Setting up anchor node

Data: Location coordinate of nodes
Result: Finding out the malicious anchor nodes Start;

Trilaterate each group of anchor nodes to a centre point and save that location;

Compare the obtained location with location reference ($M_1$);

while comparison not satisfied do
    Trilaterate all anchor nodes (individually) of the particular group (which does not satisfy the above comparison) with the neighbouring group (using the trilateration information obtained during deployment);
    Compare the obtained results with the location references ($M_2$, $M_3$, etc.);
    if a node is suspected to be malicious then
        Separate the mismatched anchors node location and save the new location in $M_N$;
    end
end
If comparison satisfied, no cheating nodes occur;
Stop;

Algorithm 2: Finding out the malicious an-chor nodes

## 3. CORRELATING WITH MAHA-LANOBIS DISTANCE

Mahalanobis distance applies posterior probability to identify the outliers. When two anchor nodes in space are demarcated by two or more associated location coordinates, Maha-lanobis distance can be used to find the dis-tance measure between the two anchor nodes. Mahalanobis distance identifies the malicious cheating nodes by comparing the location coordinates with respect to a centroid value. In our case the centroid value is the location coordinate of the trilateration point. The Mahalanobis distance function to identify the distance measure between two anchor nodes are as follows:

$$d(mahalanobis) = \sqrt{[(x_j, y_j) - (x_i, y_i)]^T * C^{-1} * [(x_j, y_j) - (x_i, y_i)]}$$

where:

$d(mahalanobis)$ is the distance between two anchor nodes,

$(x_i, y_i) \& (x_j, y_j)$ are the location coordinates of the two anchor nodes,
$C$ is the sample covariance matrix.

The variance-covariance matrix $C$ is constructed in order to gauge Mahalanobis distance,

$$C = \frac{1}{(n-1)} (x,y)^T (x,y)$$

where:

$(x, y)$ is the matrix containing the location coordinates,
$n$ is the number of nodes.

In the instance of multiple location references the variance-covariance $C$ will become as follows:

$$\begin{bmatrix} \sigma_1^2(x_i,y_i) & \rho_{12}\sigma_1(x_i,y_i)\sigma_2(x_j,y_j) \\ \rho_{12}\sigma_1(x_i,y_i)\sigma_2(x_j,y_j) & \sigma_2^2(x_j,y_j) \end{bmatrix}$$

where:

$\sigma_1^2 \& \sigma_2^2$ are the variances of the multiple location references,

$\rho_{12}\sigma_1(x_i,y_i)\sigma_2(x_j,y_j)$ is the covariance between the multiple location references.

The value of $C^{-1}$ is computed as follows:

$$\begin{bmatrix} \frac{\sigma_1^2(x_i,y_i)}{|C|} & \frac{-\rho_{12}\sigma_1(x_i,y_i)\sigma_2(x_j,y_j)}{|C|} \\ \frac{-\rho_{12}\sigma_1(x_i,y_i)\sigma_2(x_j,y_j)}{|C|} & \frac{\sigma_2^2(x_j,y_j)}{|C|} \end{bmatrix}$$

where:

$|C|$ is the variance-covariance matrix's determinant and is equal to $\sigma_1^2\sigma_2^2(1 - \rho_{12}^2)$

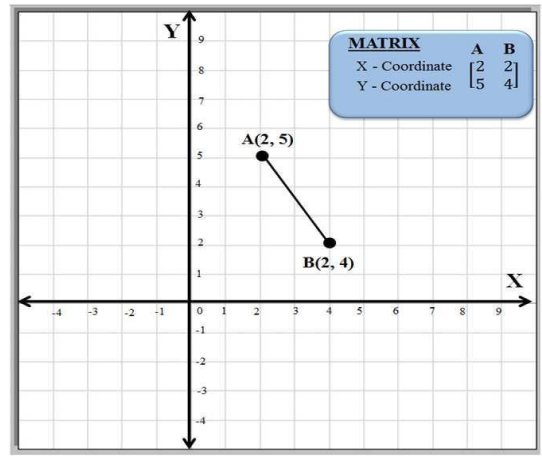

Figure 7. Transforming Location Coordinates to Matrix Representation (a).

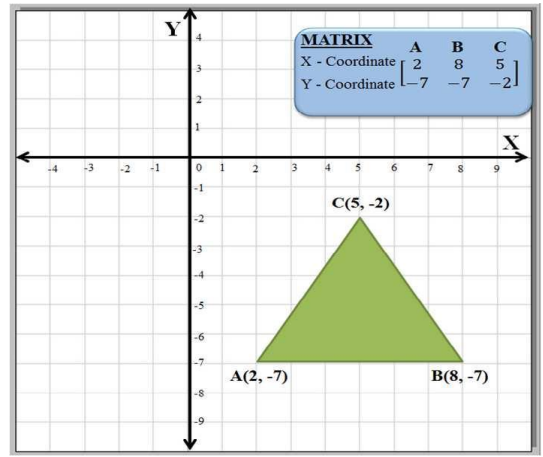

Figure 8. Transforming Location Coordinates to Matrix Representation (b).

The transformation of the location coordinates to matrix form is shown in Figure 7 and Figure 8. The Mahalanobis distance function can be modified to identify the distances from multiple location coordinates to a centroid, as follows: $d(\delta) =$

$$\sqrt{[(x_i, y_i) - (x_i, y_i)]^T * C^{-1} * [(x_i, y_i) - (x_c, y_c)]}$$

for i = 1, 2, 3, ..., n where:

$d(\delta)$ is the distance between centroid and $i^{th}$ anchor node,

$(x_i, y_i)$ is the location coordinate of the $i^{th}$ anchor node,

$(x_i, y_i)$ is the location coordinate of the centroid or trilateration point.

The new distances obtained using Mahalanobis distance, are compared using posterior proba-bility; leading to the confirmation of the anchor nodes adversity.

## 4. ASSOCIATING WITH MAXIMUM LIKELIHOOD EXPECTATION

One of the most broadly and commonly used classification technique is maximum likelihood expectation / classification. It has a good ac-ceptable result and is extensively employed and demanding algorithm.

The localization error obtained during the above mentioned algorithm is discussed in our next section. And the obtained results are compared with maximum likelihood expecta-tion method. In wireless sensor networks, all the sensor data or the sensed data are sent to a central server or aggregation point. In our scheme, the central server is made available with the location references of all the nodes in the network and MLE method is carried out with the location references available in the ag-gregation point or the central server.

Maximum likelihood Expectation is a tech-nique that is used in statistics to find the max-imum probable value from previously obtained results. The results obtained from maximum likelihood expectation can be used as the para-metric values for further experiments or simulations.

### 4.1. Probability density function

Probability density function (pdf) sorts out the required area for the random variable to occur. Consider a random sample ($x_1$, $x_2$, ..., $x_n$) from an unknown population has data vector x = ($x_1$, $x_2$, ..., $x_n$). The probability density function f (x|w) is

f (x = ($x_1$, $x_2$, ..., $x_n$)|w) =

$$f_1(x_1|w) * f_2(x_2|w) * ... * f_n(x_n|w)$$

where: x is a random sample, w is the parameter value.

Consider a scenario where n (number of trials) = 10, w = 0.4 and x = (0, 1, ..., 10), then the probability density function will be f (x|n = 10, w = 0.4) =

$$\frac{10!}{x!(10-x!)}(0.4)^x(0.6)^{(10-x)}$$

The parametric values have a large number of successive probabilities.

### 4.2. Likelihood Function

The trilateration groups are denoted as $\phi_k$, k = 1, 2, 3, ..., M where M is the number of trilateration groups. To determine the group, to which an anchor node with the current location z belongs, the conditional probabilities

$$p(\phi_k|z), k = 1, 2, 3, ..., M$$

play a crucial role. The probability $p(\phi_k|z)$ states whether $\phi_k$ is the correct trilateration group of the anchor node with the give location z. We can categorize the anchor nodes, if we know the complete set of $p(\phi_k|z)$ from decision rule

$z \in \phi_k$ if $p(\phi_k|z) > p(\phi_n|z)$ for all n =6 k    (11)

This explains that the anchor node with location z is the member of group $\phi_k$ if $p(\phi_k|z)$ is the largest probability of the set.

The desired $p(\phi_k|z)$ from the above equation and the available $p(z|\phi_k)$ from the projected training data, are correlated by Bayes theorem

$$p(\phi_k|z) = \frac{p(z|\phi_k)p(\phi_k)}{p(z)} \qquad (12)$$

where:
p($\phi_k$) is the probability that anchor nodes from group $\phi_k$ can move its location,
p(z) is the probability of finalizing an anchor node with location reference z.

$$p(z) = \sum_{k=1}^{M} p(z|\phi_k)p(\phi_k)$$

Replacing equation (12) in (11), reduces the decision rule to

$z \in \phi_k$ if $p(z|\phi_k) > p(z|\phi_n)$ for all n $\neq$ k    (13)

In equation (13), p(z) has been eliminated as a shared factor, since we don't know whether it correct location or false location. As $p(z|\phi_k)$ can be obtained from the training data, and it is plausible that the priors $p(\phi_k)$ can be esti-mated.

In order to prove mathematically, we define the discriminant function $\iota_k(z) =$

ln $p(z|\phi_k)p(\phi_k)$ = ln $p(z|\phi_k)$ + ln $p(\phi_k)$    (14)

In order to get a decision rule by substituting Eq. (14) with Eq. (13), we need a monotonic function i.e., natural logarithm. We give the decision rule as

$z \in \phi_k$ if $\iota_k(z) > \iota_j(z)$ for all j $\neq$ k      (15)

To further proceed with maximum likelihood estimation, a certain probability model is cho-sen for the trilateration group function $p(z|\phi_k)$. In our scheme, we used Gaussian distribution which is as follows: $p(z|\phi_k) =$

$(2\pi)^{-S/2}|C_i|^{-1/2} e^{\frac{-1}{2}(z-\bar{x}_i)^T C_i^{-1}(z-\bar{x}_i)}$    (16)

where: $\bar{x}$ is the mean position of the anchor node among the trilateration group $\phi_k$, $C_i$ is the covariance matrix of the trilateration group $\phi_k$, S is the N dimensional space.

To obtain the categorization function, sub-stitute Eq. (16) with (14) to get $\iota_k(z) =$
$-\frac{1}{2} S \ln 2\pi - \frac{1}{2} \ln |C_i| - \frac{1}{2}(z - \bar{x}_i)^T C_i^{-1}(z - \bar{x}_i) + \ln p(\phi_k)$ Simplifying we get, $\iota_k(z) = \ln p(\phi_k) - \frac{1}{2} \ln |C_i| - \frac{1}{2}(z - \bar{x}_i)^T C_i^{-1}(z - \bar{x}_i)$ Removing the prior probability gives us the trilateration group membership of the anchor node,

$\iota_k(z) = -\frac{1}{2}\ln |C_i| - (z - \bar{x}_i)^T C_i^{-1}(z - \bar{x}_i)$    (17)

Eq. (17) is used to identify whether the anchor node belongs to current trilateration group and reducing the localization error. If the anchor node does not belong to the group it is consid-ered deceitful.

## 5. SIMULATION AND RESULTS

Our simulation was carried out in 600m x 600m two dimensional environment. Deploy-ing the anchor node accurately is very impor-tant. First three anchor nodes were placed ran-domly and the trilateration point is found for the same. An anchor node is placed on the trilateration point attained. Any one of the first three nodes is selected and it acts as the trilateration point of the newer nodes that are going to be deployed. The above process is repeated until the final node is deployed. We deployed around 117 nodes (around 1 node for every 5m x 5m), spread randomly using the above method. Figure 9 shows the deployment of the anchor nodes in our scenario.

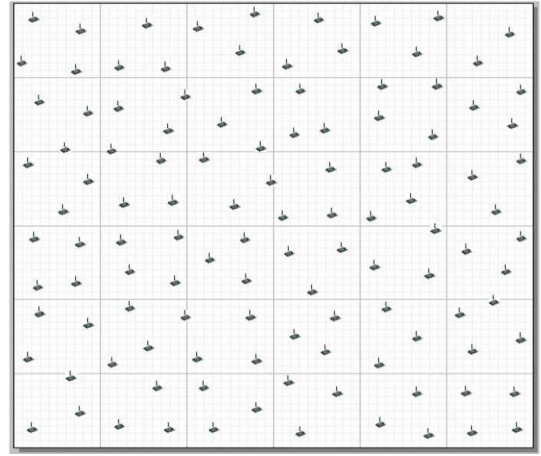

Figure 9. Deployment of Anchor Nodes.

### 5.1. Experiment using trilateration tech-nique

Few anchor nodes were compromised (mak-ing it transmit false information regarding its current location) randomly and the malicious anchor nodes were found out using trilatera-tion technique. The localization error tran-spired while localizing the malicious anchor nodes from random samples, were noted down. Each simulation was carried out for 50 times and the mean error was considered. Figure 10 shows the mean error in location discovery and Figure 11 shows the time taken to locate the malicious anchor nodes during simulation.

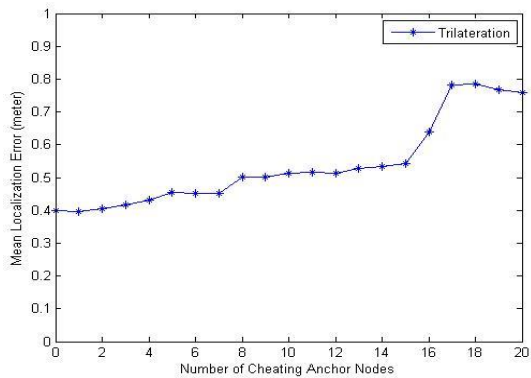

Figure 10. Mean Localization Error While using Trilateration.

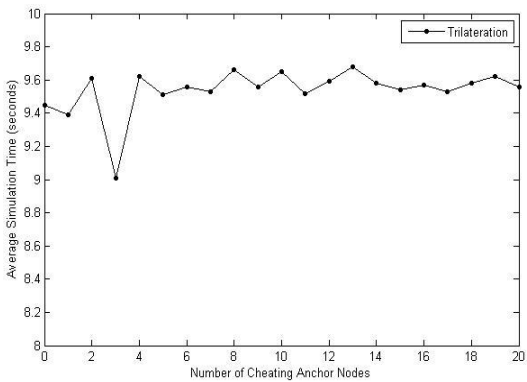

Figure 11. Average Time for Simulation.

### 5.2. Comparing with Mahalanobis dis-tance

The central server or aggregation point has a list of initial location references of the anchor nodes. The false location of the malicious anchor nodes obtained, were compared with the results obtained from Mahalanobis distance. Comparing the results obtained, reduced the error in location discovery. Figure 12 shows the mean error in locating malicious anchor nodes.

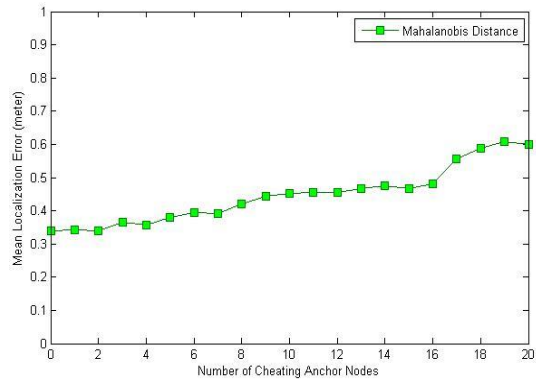

Figure 12. Mean Error after Comparing with Mahalanobis Distance.

### 5.3. Comparing with Maximum Likeli-hood Expectation

Maximum likelihood function has a list of ini-tial location references of the anchor nodes. The false location of the malicious anchor nodes obtained, were compared with the re-sults obtained from maximum likelihood func-tion. Comparing the results obtained, reduced the error in location discovery. Figure 13 shows the mean error in locating malicious anchor nodes while using maximum likelihood expec-tation. Figure 14 shows the comparison of the three results, trilateration, trilateration with Mahalanobis Distance and trilateration with MLE. Finally the information about the mali-cious anchor node is conveyed to all the nodes other than the infected nodes, and the routing table is updated by confiscating the malicious anchor node.

### 6. DISCUSSION AND FUTURE EVENTS

Malicious anchor nodes will constantly hin-der genuine and appropriate localization. Our scheme was carried out using fixed sensor nodes and the attack has a permanent consequence in the sensor node. Reducing the localization er-ror and endorsing the malicious anchor node were implemented successfully in this paper.

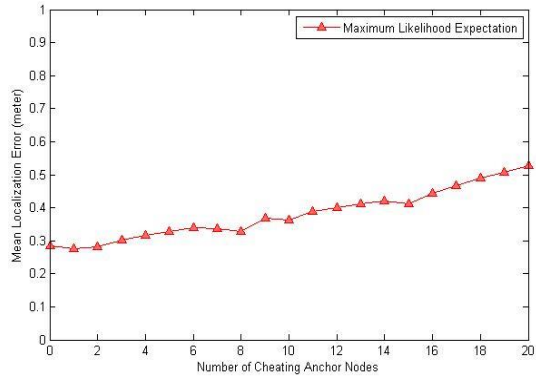

Figure 13. *Mean Error after Comparing with Maximum Likelihood Function.*

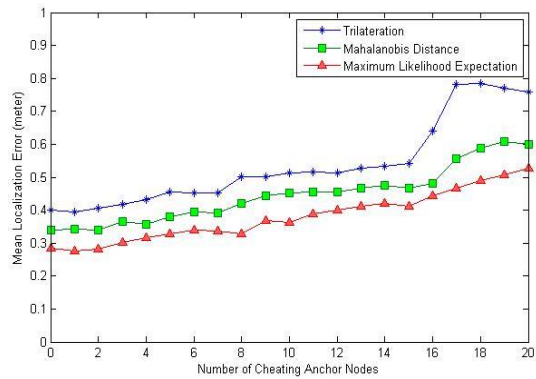

Figure 14. *Result Comparison.*

We have proposed a novel scheme using maximum likelihood and trilateration technique to identify malicious anchor nodes. The error can be increased if hindrance, interferences, and attenuation caused by signal fading, and noise are additional. Our scheme can also be mod-elled to overcome such disturbances by using some statistical distributions like Rayleigh or Rician distributions [34]. Our algorithm per-formed consistently for different topologies.

Our scheme can be extended for mobile sensor node with an intermittent attack type. Our framework can be extended to acoustic and ultra-wideband (UWB) technology. Using en-ergy efficiency as a benchmark is quite chal-lenging. Our algorithm was implemented in 2-D plane and can be extended to 3-D plane also.

## 7. CONCLUSIONS

For smart environments, security plays a very essential part. In this paper, we discussed about localizing malicious anchor nodes in a secured manner, using trilateration technique and comparing the results obtained with max-imum likelihood expectation and Mahalanobis distance. By both the techniques way we were able to reduce the error attained during local-ization. However, maximum likelihood expec-tation outperformed Mahalanobis distance in perceiving cheating beacon nodes. By using maximum likelihood expectation and Maha-lanobis distance we can obtain consistent and proficient results. Our results show that as the malicious anchor nodes increases, the simula-tion time and error obtained during location discovery slightly increases. The accuracy ob-tained in our work can be used as assistance in some wireless applications. Some imminent events for further research have been discussed.

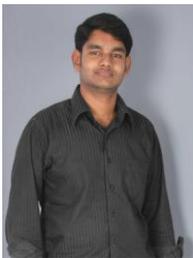
Jeril Kuriakose received the B Tech degree from Jeppiaar Engineering College, India, in 2010, and M Tech degree from University of Mysore, India, in 2012, all in Information Tech-nology. He is currently pursuing Ph.D in Manipal University Jaipur, India. His research inte-rests include Mobile Computing, Mobile Ad Hoc Network, Network and Information Security and Scientific Computing.

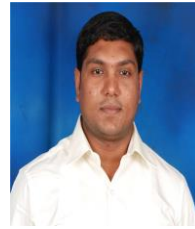
Amruth V received B Tech Degree from Coorg Institute of Technology, India and MTech Degree from University of Mysore, India, in 2009 and 2012, respectively, all in Infor-mation Technology. At present he is working as assistant prof-essor in Bearys Institute of Technology, Manga-lore, India, in the Department of Information Science and Engineering. His research areas in-clude Remote Sensing, Algorithms, Network and Information Security and Wireless Networking.

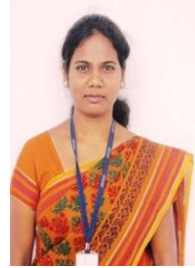
Swathy Nandhini N com-pleted UG degree in KSR College of Engineering and PG degree in Varuvan Vadivelan Institute of Technology, Anna University, in India, in 2010 and 2012, respectively. Her area of interest includes Cryptography and Network Security, Comput-er Networks, Operating Systems and Web Tech-nology.

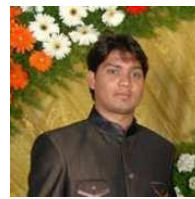
Abhilash V is an Electrical an Engineer. He completed his bachelors from PES College of Engineering, Mandya. His re-search areas are Power Systems and Control Systems.